\documentclass{pasj00}

\newcommand{\chg}[1]{#1}

\begin{document}
\SetRunningHead{Miyaji et al.}{ASCA Observation of Kaz 102}
\Received{}
\Accepted{}

\title{{\it ASCA} observation of Unusually X-ray Hard Radio Quiet 
 QSO Kaz 102}

\author{Takamitsu  \textsc{Miyaji}\altaffilmark{1},
        Yoshitaka \textsc{Ishisaki}\altaffilmark{2},
	Yoshihiro \textsc{Ueda}\altaffilmark{3},
	Yasushi \textsc{Ogasaka}\altaffilmark{4},
	Hisamitsu \textsc{Awaki}\altaffilmark{5},\\
	Kiyoshi \textsc{Hayashida}\altaffilmark{6}
} %

\altaffiltext{1}{Department of Physics, Carnegie Mellon University, 
     Pittsburgh, PA 15213, USA}
\email{miyaji@cmu.edu}
\altaffiltext{2}{Department of Physics, Tokyo Metropolitan University,
   Minami Osawa 1-1, Hachioji, Tokyo 192-0397}
\email{ishisaki@phys.metro-u.ac.jp}
\altaffiltext{3}{ISAS, 3-1-1 Yoshinodai, Sagamihara, Kanagawa 229-8510}
\email{ueda@astro.isas.ac.jp}
\altaffiltext{4}{Department of Astrophysics,
       Nagoya University, Nagoya, Aichu 464-01}
\email{ogasaka@u.phys.nagoya-u.ac.jp}
\altaffiltext{5}{Department of Physics, Ehime University,
    Matsuyama, Ehime 790-8577}
\email{awaki@sgr.phys.sci.ehime-u.ac.jp}
\altaffiltext{6}{Department of Astrophysics, 
 Osaka University, Matikane, Toyonaka, Osaka 560-0043} 
\email{hayasida@ess.sci.osaka-u.ac.jp}


%


\maketitle

\begin{abstract}
We have observed a radio-quiet QSO Kaz 102 (z=0.136) with 
{\it ASCA} as a part of our program of complete spectral 
characterization of hard X-ray selected AGNs. We found that Kaz 102 
shows unusual spectral properties. A simple power-law with absorption
in our galaxy gave a satisfactory description of the spectrum. 
However, it showed a very hard photon index of $\Gamma \sim 1.0$ 
with no sign of deep absorption or a prominent spectral feature. 
We further explored the Compton reflection with Fe K$\alpha$ line 
and warm absorber models for hardening the spectra. Both gave 
statistically satisfactory fits. However, the Compton reflection
model requires a very low metal abundance (0.03-0.07 in solar units).
The warm absorber model with no direct component is preferred and 
gave a very high ionization parameter $\xi \sim 10^{2.3}$. If this is
the case, the values of $\xi$, warm absorber column density, and
variability over $\sim 10$ years may suggest that the warm absorber
resides in the broad-line region and crosses the line of sight to
the central X-ray source.  
   
\end{abstract}

\section{Introduction}
\label{sec:intr}

 The underlying intrinsic X-ray continua of AGNs without
apparent deep absorption usually show a power-law form with a typical 
photon index of $\Gamma \sim 1.7-2.0$.  There is some dispersion in the 
spectral indices, however. For example, Narrow-line Seyfert 1 galaxies 
(NLS1) are known to have intrinsically softer spectra, with up to 
$\Gamma\approx 2.5$ (e.g. \cite{leighly}).   AGNs with apparently much 
harder spectra usually arises from some reprocessing, i.e., absorption 
(by neutral or partially ionized gas) and/or Compton-reflection. These 
reprocessing leaves some characteristic features, which are sometimes
hard to recognize in X-ray spectra with low signal-to-noise
ratio.  

 As a part of our study of a complete determination of spectral
properties of a hard X-ray selected bright AGNs using {\it ASCA}
and {\it XMM-Newton} \citep{mrs01}, the primary goal of which is to 
serve as a bright-end constraint to hard X-ray luminosity function, 
evolution, and X-ray AGN population synthesis studies (e.g. Ueda et al. 
in preparation), we have observed a radio-quiet QSO (RQQ) Kaz 102 
(a.k.a. 1803+676) with {\it ASCA}. The results of the study of the whole 
sample will be reported elsewhere (Miyaji et al. in preparation). 

 Kaz 102 is an optically-bright RQQ at $z=0.136$ with broad
lines (H$\alpha$ FWHM $\approx 5000\;{\rm km\,s^{-1}}$ \citep{treves,dediego}  
and was already recognized as a very hard X-ray source with a photon 
index of $\Gamma\sim 0.8^{+0.6}_{-0.4}$, when it was observed by the 
{\em Einstein Observatory} in 1979 \citep{wilkes}.  
However, during the {\em ROSAT} All-Sky Survey (RASS) in July 1990-
January 1991, the spectral index in the {\it ROSAT} band was much 
softer ($\Gamma \sim 2.2$), which was typical of an RQQ in the 
{\rm ROSAT} band \citep{ciliegi,treves}. 
In this letter, we report the unusual spectral properties of this 
object during our {\it ASCA} observation. Luminosities are calculated 
using $H_{\rm 0}=65$ km\,s$^{-1}$Mpc$^{-1}$, $\Omega_{\rm m}=0.3$ and 
$\Omega_{\Lambda}=0.7$. 

\section{ASCA Observation and Analysis}

\subsection{Observation}

 The ASCA observation was made during the ASCA AO7 period as a
part of our larger program (see above). The log of observation of Kaz 102 
is shown in Table \ref{tab:log}. Data reduction and extraction of events 
have been made using {\bf FTOOLS 5.0} or later and spectral fittings have 
been made using {\bf XSPEC 11.2}
\footnote{http://heasarc.gsfc.nasa.gov/docs/xanadu/xspec/}. 

\begin{table}
  \caption{Log of {\it ASCA} observation of Kaz 102}\label{tab:log}
  \begin{center}
    \begin{tabular}{lllll}
      \hline\hline
      Sequence &Obs. Date    & Exposure  & SIS/GIS Modes \\\hline
      77015000 &Sep. 5, 1999 & 20 ks     & 1CCD-Faint/PH  \\
      \hline
    \end{tabular}
  \end{center}
\end{table}

 We found no significant variability of X-ray flux during the {\it ASCA} 
observation.

\subsection{Spectral Extraction and Analysis}
\label{sec:spec}

\begin{figure}
  \begin{center}
    \FigureFile(80mm,80mm){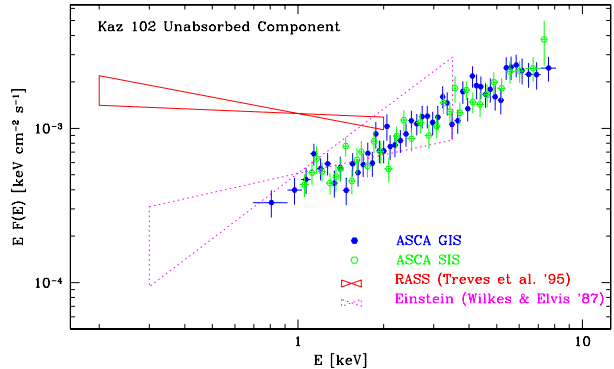}
  \end{center}
  \caption{The unfolded $E\;F(E)$ GIS and SIS spectra are
  shown with the results of the power-law fits of the ROSAT
  All-sky Survey \citep{treves} and Einstein \citep{wilkes} spectra.}
  \label{fig:efe}
\end{figure}

 The SIS and GIS pulse-height spectra have been created from the events 
screened with the standard screening criteria. Background has been
accumulated from the off-source area of the same observation for
the SIS spectra. For the GIS spectra, we have used backgrounds from both 
the off-source area of the same observation and the standard-screened
blank sky observation with a similar $N_{\rm H}$ value to that of our 
target location (using the FTOOLS task {\bf mkgisbgd}). There was very 
little difference in the  GIS results between these two background 
estimation methods, and thus we present the results with the background 
from the off-source area of the same observation. Spectra from two 
SISs (SIS0/SIS1) and the two GISs (GIS2/GIS3) have been co-added 
separately and the co-added SIS and GIS spectra have been analyzed jointly. 
We have made spectral analysis using pulse height channels corresponding 
to 1-7 keV and 0.7-9 keV for SIS and GIS respectively. 
{We ignored SIS channels below 1 keV, where the responses are
affected by radiation damage and there are calibration uncertainties
associated with it for observations during late stages of the mission. 
In any event, because of the limited statistics, the impact 
of this effect, corresponding to a spurious excess absorption of 
$N_{\rm H}\approx 5\times 10^{20}$ ${\rm cm^{-2}}$ \citep{yaq_calrep},
is not significant, as demonstrated by the upper limit to the extra 
absorption over the Galactic value (Sec. \ref{sec:pl}).}
All errors and upper limits correspond to a 90\% confindence level 
($\Delta\;\chi^2=2.7$). The unfolded GIS and SIS spectra corrected for 
the Galactic absorption are shown in Fig. \ref{fig:efe} in the $E\;F(E)$ 
space. The spectra of previous X-ray observations, from
{\it ROSAT} All-Sky Survey \citep{treves} and {\it Einstein}
are also shown. 

\subsubsection{Single Power-Law and Galactic Absorption}\label{sec:pl}

\begin{figure}
  \begin{center}
    \FigureFile(80mm,150mm){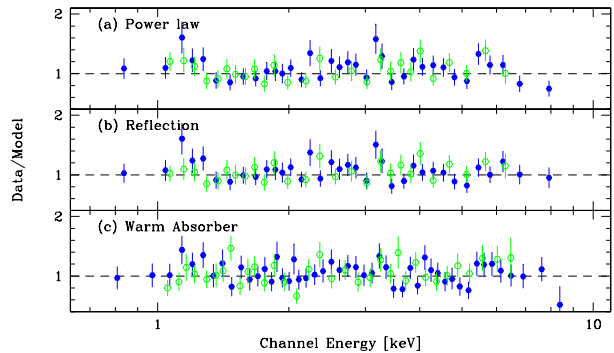}
  \end{center}
  \caption{The data/model ratios of the three spectral fits 
  discussed in Sect. \ref{sec:spec} are shown with 1$\sigma$ 
  errors.}
  \label{fig:resid}
\end{figure}

 Firstly we have tried to fit the {\it ASCA} GIS and SIS spectra with
a single power-law. A Galactic absorption component, with the column
density fixed to $N_{\rm H}=4.6\times 10^{20}$ ${\rm [cm^{-2}]}$, 
is included in all the spectral fittings in this and following 
subsections. The spectral slope was a joint parameter and the global 
normalizations were separate parameters. When additional neutral absorber 
located at the redshift of the object ($z=0.136$) is added, we found 
a 90\% upper limit of $N_{\rm H}^{\rm z}< 1.0\times 10^{21}$ 
${\rm cm^{-2}}$). The result of the fit is under column (1) of  
Table \ref{tab:fit}, giving a photon spectral index of 
$\Gamma=1.02\pm .06$ with 
$\chi^2/\nu=1.01$. The 2-10 keV luminosity at the object's rest
frame is $L_{\rm x}\approx 2.3\times 10^{44}$ erg\,s$^{-1}$, which 
is the luminosity of a moderately powerful QSO. The residuals from the 
best-fit power-law model is shown in Fig. \ref{fig:resid}(a). 
 
 As seen in  Fig. \ref{fig:resid}(a), we see a weak edge-feature at the
object's rest frame of $E_{\rm ed}=1.5\pm $0.1 keV and 
$0.2<\tau_{\rm ed}<0.7$ (90\% confidence). \chg{A similar feature 
is seen in the {\it ASCA} spectrum of NGC 4395 \citep{iwas_4395}.} 
Adding the edge improved the fit by $\Delta \chi^2=16$ with two 
additional degree of freedom (F-test probability of $5\times 10^{-3}$).   

\subsubsection{Compton Reflection Model}  \label{sec:refl}

 We have tried to fit the spectrum of this AGN with a canonical 
$\Gamma=1.9$ intrinsic power-law and an additional Compton
reflection component. We tried {\bf XSPEC} models for both the ionized 
({\em pexriv}) and neutral ({\em pexrav}) reflectors \citep{pexrav}.
{The best-fit neutral reflector ({\em pexrav}) model with 
$\cos\;i =0.45$ (fixed) gave $\chi^2/\nu=0.95$, where $\pi/2-i$ is 
the angle between the line of sight and the plane of the reflector surface.} 
When a narrow Gaussian (width is fixed at 0.01 keV) is added at the redshifted 
Fe K$\alpha$ line energy of 5.6 keV (fixed), the fit has marginally 
improved with $\Delta \chi^2 =3.6$ (F-test probability of 6\%), while 
the values of other parameters did not change dramatically. The fit 
result in this case is shown in Table \ref{tab:fit} column (2).  
Equally acceptable fits have been obtained for  $\cos\;i$ values 
of 0.1 and 0.8 with almost the same $\chi^2$ values, giving 
{\it rel\_refl}=6.5 and 29 respectively. 

{In the {\em pexriv} fit, where the refletor temperature (T)
and the ionization parameter ($\xi$, in units of ${\rm [erg\;s^{-1}\;cm]}$ 
defined by \citet{done92}) are assumed to satisfy  the equilibrium 
condition (see below), we only obtain an upper limit $\xi < 100$ 
($T<4\times 10^5{\rm ^\circ K}$) for $\cos\;i =0.45$.  
Varying the values of $\xi$ within this limit is compensated by the 
variation of {\it rel\_refl} parameter and the normalization, where 
their best-fit values become {\it rel\_refl}$\approx 21$, $S_{\rm x12}=0.5$ 
(direct component only) and abundance=0.1 at this upper limit.}  

 The range of {\it rel\_refl} indicates that one requires 
at least several times stronger source of incident radiation than that 
is seen as the direct component in order to produce the reflection 
component. This could be because some blocking material 
partially covers our line of sight to the original power-law source 
but the reflectors can be seen. Alternatively, the original source may 
have become fainter, while the reflected X-rays arrive later with 
extra paths.  

  The 90\% confidence range of the Fe K$\alpha$ equivalent 
width (EW) from the fit was 10-290 eV. The abundance parameter
of {\em pexrav} is constrained to be very low, with a 90\% confidence  
range of 0.03-0.07 in solar units. {With solar abundance, 
the data fail to fit with concavity around photon energies at 2-4 keV 
from a collection of numerous edges. The constraint from the single Fe K 
edge play a minor role because of the limited photon statistics at 
$E\gtrsim 6$ keV. At the position of the redshifted Fe K$\alpha$ emission 
line (5.6 keV), the reflection component dominates and 
Fig. 2 of \citet{balla02} can be used to find constraints from the 
Fe K$\alpha$ EW. At this low abundance, the EW of $\sim 300$ eV or 
lower is consistent with for all values of $\xi$. Conversely, this 
EW range corresponds to an Fe abundance of $\lesssim 0.2$ for a 
neutral reflector, but for our upper limit of $\xi$ ($\sim 100$) in 
equilibrium, the constraint becomes looser (Fe abundance $\lesssim$ 1).} 
Thus we cannot exclude the possibility that the X-ray emission 
of this object is reflection dominated from the spectral fit alone.  
 
\subsubsection{Warm Absorber Model}\label{sec:warm}
  
\begin{figure}
  \begin{center}
    \FigureFile(80mm,80mm){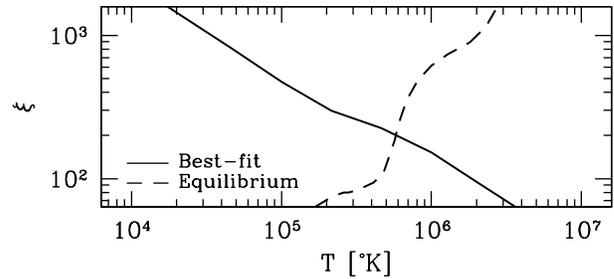}
  \end{center}
  \caption{Solid curve shows the best-fit ionization parameter 
   $\xi$ for given absorber temperatures $T$. The dashed curve 
   show the condition where the absorbing gas is in equilibrium
   \citep{reynolds}.}    
  \label{fig:xi_t}
\end{figure}

 We have also considered a model with  $\Gamma=1.9$ intrinsic power-law 
(fixed), and a warm absorber, using the XSPEC model 
{\em absori} \citep{done92,absori_z}. One of the motivations of considering
this is the existence of the absorption edge as described 
in Sect.\ref{sec:pl}. \chg{The {\em absori} model does not include 
the solution for local radiative transfer and thermal-ionization 
balance, and it requests the absorber's temperature $T$ and  
ionization parameter $\xi$ as independent parameters. We have 
approximately solved for $\xi$ and $T$ as follows. 
These two parameters are strongly correlated and cannot be determined 
independently by the spectral fit.  Thus we have first determined the 
best-fit $\xi$ as a function of $T$. This is plotted in 
Fig. \ref{fig:xi_t} as a solid line. We then 
overplot the $\xi$--$T$ relation (dashed line) in the thermal-ionization 
equilibrium case from Fig. 4 of \citet{reynolds} for the $\Gamma=1.8$ 
ionization source, which is the closest to our assumption.} These two curves
give a solution of  $(\xi, T)=(200, 6\times 10^5\;{\rm ^\circ K}$).  
   
 The fit result is shown under column (3) of Table \ref{tab:fit}, 
where $\xi$ and $T$ are jointly varied along the equilibrium line during 
the fit and error search. In this table,  $N_{\rm H22}^{\rm z}$ 
represents the column density of the warm absorber and the abundance 
is the iron abundance in solar units, and $\xi$ is the ionization parameter. 
Indeed the best-fit warm absorber model gave a blend of edge features 
near $E \sim 1.5$ keV and the $\chi^2/\nu$ value has been improved 
compared with the single power-law case partially because of this. 
A caveat is that the errors in these parameters represent only 
statistical errors of the fit under restrictive assumptions and 
should be treated accordingly. The analysis shows that if the warm 
absorber model is the case, the line of sight to
the X-ray source is totally  covered by a highly ionized 
($\log_{\rm 10}\xi\sim$ 2.3) slab of gas with a column density of 
$\approx 3\times 10^{23}$ ${\rm cm^{-2}}$.

\begin{table}
\caption{Results of the Spectral Fits}\label{tab:fit}
\begin{center}
 \begin{tabular}{lcccc}
\hline\hline
Parameter              & PL            & Reflection & Warm abs.\\
                       & (1)           & (2)    &  (3)    \\\hline
$\Gamma$               & $1.02\pm .06$ &  1.9 (fixed)   &  1.9 (fixed)   \\
$N_{\rm H22}^{\rm z\;\;a)}$  & $ <0.1$     & \ldots  & $34\pm 11$\\
$S_{\rm x12}^{\rm PL\;b)}$ & $4.8\pm .3$   & $.9\pm .1$ & $4.5\pm .4$\\
$\log_{\rm 10}\xi ^{\rm \;c)}$ & \ldots     & $<2.0$ & $2.3\pm .1$\\
$T\;{\rm [10^5\;^\circ K]^{\rm \;c)}}$& \ldots & $<4$ & $5.8\pm .4$\\     
abundance          & \ldots        & .05$\pm$.02 & .2$\pm$.1\\ 
rel\_refl          & \ldots        & 9.2$\pm$1.3 & \ldots    \\
EW(Fe)$^{\rm \;\; d)}$   & \ldots        & 150$\pm$140 & \ldots \\              
$\chi^2/\nu$            & 157./156& 145./155 & 150./155\\\hline
\end{tabular}
\end{center}
Notes: Galactic absorption is included in all models. All upper limits
have been fixed to 0 for the determination and error search of other
parameters. $^{\rm a)}$ In units of 10$^{22}$  ${\rm [cm^{-2}]}$. 
This represents column densities of neutral and warm absorbers for 
columns (1) and (3) respectively.  $^{\rm b)}$ 
The {\it ASCA} GIS 2-10 keV flux 
in units of  10$^{-12}$  ${\rm [erg\;s^{-1}\;cm^{-2}]}$. This 
represents the fluxes of the intrinsic (before absorption) for (1) 
and (3) and of the direct component only for (2) respectively.
$^{\rm c)}$ T and $\xi$ are linked assuming the equilibrium condition. 
See text.
$^{\rm d)}$ The equivalent width of the Fe K$\alpha$ line in eV 
in the object's frame.   
\end{table}

\section{Discussion}\label{sec:disc}

\begin{figure}
  \begin{center}
    \FigureFile(80mm,40mm){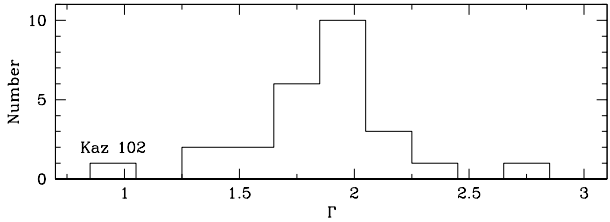}
  \end{center}
  \caption{The distribution of the photon index $\Gamma$ of
  {\em unabsorbed} AGNs in our hard X-ray selected 
  sample (\cite{mrs01}; Miyaji et al. in prep.) with Kaz 102
  labeled.}    
  \label{fig:gamhis}
\end{figure}

 Kaz 102 has an apparently unusual spectral property.  
Fig. \ref{fig:gamhis} shows the histogram of the photon 
indices ($\Gamma$) of 28 of the $\sim 50$ AGNs in our sample 
(see Sect. \ref{sec:intr}) which seem unabsorbed, i.e., those
which have intrinsic absorption of $N_{\rm H}<10^{21} {\rm cm^{-2}}$
in the simple absorbed power-law fits.  Fig. \ref{fig:gamhis} shows 
the unusually small spectral index among these ``unabsorbed'' AGNs,
and also harder than the range of spectral indices of 26 RQQs 
observed by \citet{george00}.  It is also worth noting that 
this spectral index is difficult to produce with the current standard 
theory of AGN hard X-ray continuum involving inverse-Compton scattering 
by a hot corona (e.g. \cite{haardt93}). Thus this object deserves an 
individual attention.  
   
\begin{figure}
  \begin{center}
    \FigureFile(80mm,150mm){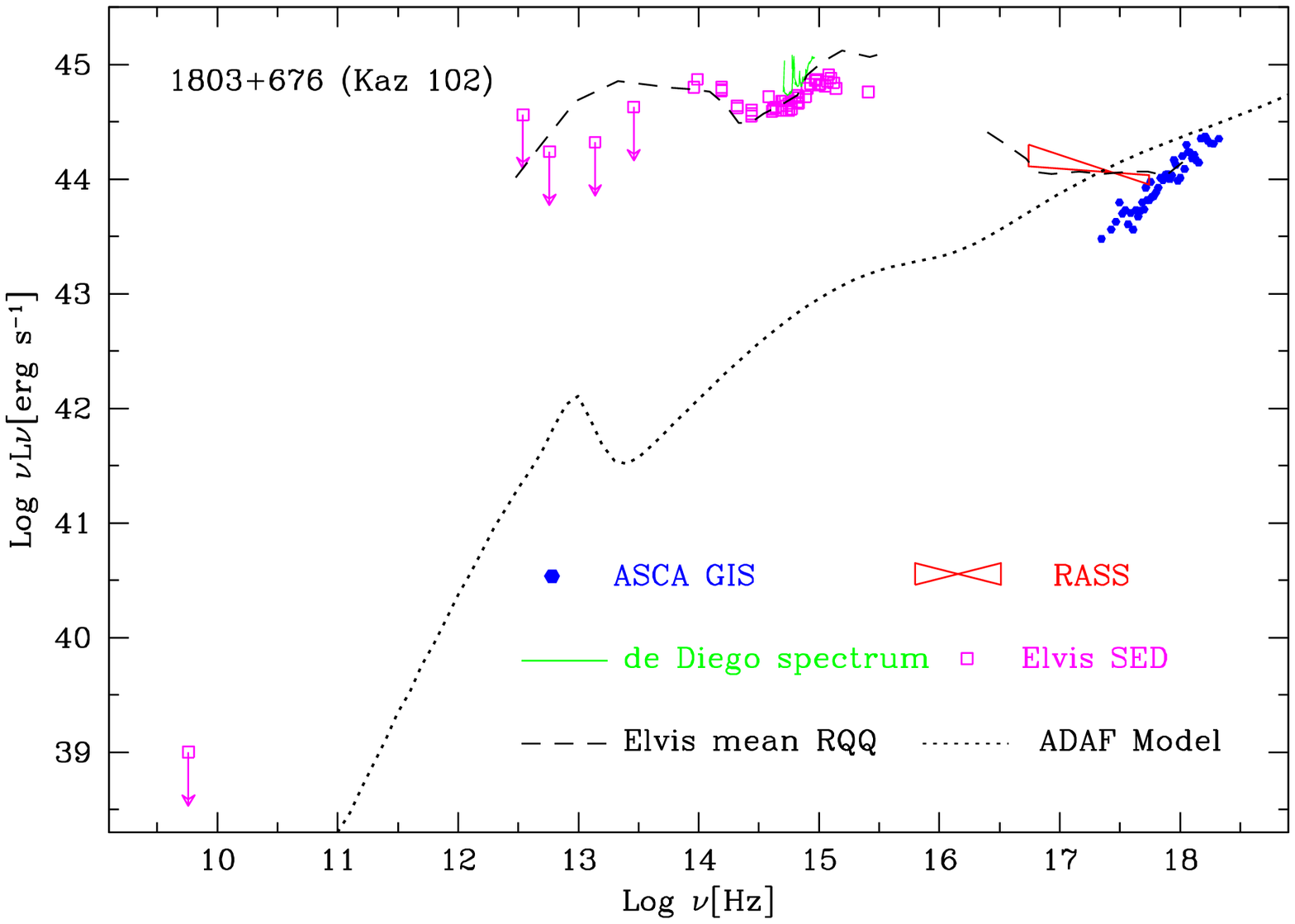}
  \end{center}
  \caption{The infrared-Xray spectral energy distribution of Kaz 102 is 
  plotted. The radio-optical photometries are from \citet{elvis94}, optical
  spectrum is from \citet{dediego}, and X-ray data points are from 
  Fig. \ref{fig:efe}. The mean RQQ SED from \citet{elvis94} is also
  shown. The SED of an ADAF model is shown for  
  $M_{\rm BH}=10^8\;M_\odot, \dot{m}=\dot{m}_{\rm crit}$ 
  (Di Matteo, priv. comm) is also plotted for comparison.  
  All symbols and line styles are labeled in the figure.}    
  \label{fig:sed}
\end{figure}

To understand the nature of this object better, we plot the infrared to 
hard X-ray spectral energy distribution (SED) of Kaz 102 in 
Fig. \ref{fig:sed}. For a comparison, the mean SED of radio-quiet 
QSOs (RQQ) \citep{elvis94} normalized to the luminosity of this object 
at around $\lambda=6000$ \AA ~ is overplotted. We see that the 
{\it ROSAT}  spectrum is well on the mean RQQ SED.   

 From the three spectral fits shown in Sect. \ref{sec:spec}, we found 
statistically satisfactory fits in any case. We consider the 
reflection-dominated picture unlikely because the fit 
requires unusually low heavy element abundance of less than
0.1 solar, which is much less than clusters/groups of galaxies and
is not likely for gas in galaxies. We have also considered the 
possibility of accretion flow (ADAF) with a mass accretion rate 
$\dot{m}$ close to the critical value $\dot{m}_{\rm crit}$, above 
which ADAF solutions do not exist, as the origin of the hard X-ray 
emission. The X-ray portion of the calculated SEDs of the ADAF spectrum 
from Di Matteo (priv. comm) based on the model described in 
\authorcite{dimatteo99} (\yearcite{dimatteo99,dimatteo03}) are 
overplotted in Fig. \ref{fig:sed} for  $M_{\rm BH}=10^8\,M_\odot$ and  
$\dot{m}=\dot{m}_{\rm crit}$.  These parameters are chosen to
give approximately the luminosity of Kaz 102 at $E\sim 8$ keV. 
We see that the X-ray portion of this ADAF model ($\Gamma \sim 1.4$) 
is not hard enough to explain our {\it ASCA} X-ray spectrum of Kaz 102
and also fail to explain the optical-UV portion of the SED
by a large factor. This is essentially due to the fact that
we see the bing blue bump (BBB) in the UV region, which does not
exist in the ADAF model. 

 The existence of the sign of an edge feature gives preference to the 
warm absorber explanation of this object. The best-fit abundance in 
the warm absorber model using {\em absori} is still sub-solar 
(0.1-0.3), \chg{which may be too low for a massive system hosting 
such a luminous QSO. However, it may be is possible to explain the 
X-ray spectrum with higher abundance gas in more complex models, e.g.
involving multiple absorber models and/or models involving 
both the Compton reflection and warm absorber, since having 
multiple components tend to smooth out edge features, giving a 
lower abundance when fit with a simple model.} 
Our data cannot make constraints on these complex models and should 
be a topic of future X-ray observations. 
 
 An interesting observation is that we see a soft component at $E<1.5$ keV 
during the {\it ROSAT} observation in 1990, but not during the {\it Einstein} 
(1979) and {\it ASCA} (1999) observations. The difference of the {\it ROSAT} 
and {\it ASCA} spectral slopes in the overlapping energy range far exceeds 
the reported discrepancy in fitted spectral slopes between these instruments 
during simultaneous observations of NGC 5548, which suggested 
spectral response calibration errors in either or both instruments 
\citep{iwas_rosasca}. {Thus the variability of the soft
component cannot be attributed to the cross-calibration problem. 
Such a variability of the soft component have previously been 
reported in a number of AGNs including 1H 0419-577 \citep{guai98,page02}
and REJ2248-511 \citep{puch95}.} The soft excess component observed by 
{\it ROSAT} is not the high energy end of the BBB, which is thought 
to come from the multi-temperature blackbody emission of the accretion 
disk. Integrating mid-infrared to ultraviolet SED of Kaz 102 gives 
a lower limit to its bolometric luminosity 
$L_{\rm bol}\gtrsim 2\times 10^{45}$ erg\,s$^{-2}$, corresponding to 
an Eddington mass of $\gtrsim 3\times 10^7$ $M_\odot$. 
The temperature of the inner accretion disk 
(assuming non-rotating) around a black hole with this mass is as low as 
$kT \lesssim 0.02$ keV (e.g. Eq. (12) of \cite{makishima}), and thus 
can very little contribute to the X-ray emission at $E \sim 0.5-1$ keV.
{\citet{page02} explains the soft variability of 1H 0419-577
as due to the cooling of comptonizing electrons in the framework
of two-temperature coronae and this may also be the case for Kaz 102.}

 The second interpretation is that the line of sight to 
the non-thermal continuum is sometimes blocked by the crossing 
warm absorber cloud.  In the framework of the crossing warm absorber 
picture, we can put order-of-magnitude constraints on the nature of 
the warm absorbing cloud using the value of 
$\xi \equiv L_{*}/nr^2\approx 200$, where 
$L_{\rm *}\approx 2\times 10^{45} {\rm erg\;s^{-1}}$ is the 5eV-300 keV 
luminosity of the central source irradiating the cloud, estimated from 
the extrapolation of the unabsorbed power-law component. 
From the $N_{\rm H}$ of the warm absorber and $\xi$, we can calculate 
hydrogen+ion density $n$ and physical depth $\Delta r\approx N_{\rm H}/n$ 
of the cloud for a given distance $r$ from the X-ray source. Furthermore, 
assuming that the central blackhole of $M_{\rm BH}\sim 10^8 M_\odot$ dominate 
the gravitational force in the region of interest ($r<$ a few pc) and 
cloud dimensions in depth and width are similar, its circular velocity  
($v_{\rm cir}=GM_{\rm BH}/r$) and the cloud crossing time 
($t_{\rm cr}=\Delta r/v_{\rm cir}$) can be estimated. Table 
\ref{tab:cloud} shows the results of the calculation. {A geometrical 
constraint $\Delta r<r$ gives $r<10^{19}$ cm.} Taking the crossing warm 
absorber cloud picture and assuming that $t_{\rm cr}$ is about the 
timescale of the variability ($\lesssim$ several years), the absorbing 
cloud is most likely to be located at $r\lesssim 10^{18}$ cm 
($\approx 0.3$ pc) and the value of $v_{\rm cir}\gtrsim 1000$ 
${\rm km\,s^{-1}}$ there suggests that it is in the broad-line region. 

\begin{table}
  \caption{Warm Absorber Cloud Parameters}\label{tab:cloud}
  \begin{center}
    \begin{tabular}{ccccc}
      \hline\hline  
    $r\;{\rm [cm]} $ & $n\;{\rm [cm^{-3}]}$ & $\delta r\;{\rm [cm]}$ & 
       $v_{\rm cir}\;{\rm [km\;s^{-1}]}$ & $t_{\rm cr}$ \\
    \hline
    10$^{19}$ & $10^{5}$ & $10^{18.5}$ & 
            $400$ & $2000$ yr. \\
    10$^{18}$ & $10^{7}$ & $10^{16.5}$ & 
            $1000$ & $10$ yr. \\
    10$^{17}$ & $10^{9}$ & $10^{14.5}$ & 
          $4000$ & $10$ days. \\
    \hline
    \end{tabular}
  \end{center}
\end{table}

 The third explanation is the change of ionization state of the warm 
absorber. If the central source was twice as 
luminous, i.e. $\xi$ increased to 400 (the absorber temperature 
increases accordingly), the absorber becomes more transparent at 
$\lesssim 2$ keV, reproducing the steep spectrum of the {\it ROSAT} 
observation. However, in the case of Kaz 102, this is not a likely 
explanation, because the entire spectrum had to be twice as luminous 
when {\it ROSAT} observed the soft excess component. It might be 
possible, however, that {\it ROSAT} observation was made just after 
the irradiating source had become fainter, but the absorber retains 
the previous ionization state. This is not likely the case unless the 
cloud is located just at the geometrical constraint threshold 
of $r\approx 10^{19}$ cm with $n\approx 10^5$ cm$^{-3}$ 
(Table \ref{tab:cloud}). The reason is that \citet{treves} detected
no significant variability of the soft X-ray flux ($\lesssim 10$\% if any) 
during the repeated {\it ROSAT} All-Sky Survey scans through 
Kaz 102 over $\sim 180$ days, except for a sign of two brief dips. 
And the recombination timescale estimated by Eq. (2) of \citet{reynolds} 
is $\approx 10^{12}/n$ s, which become comparable or longer than this 
timescale only if density is $n\lesssim 10^5$ cm$^{-3}$, hence
$r\gtrsim 10^{19}$ cm. 

 In order to discriminate between the discussed possibilities 
of the origins of the hard X-ray spectrum and the variability
of the soft component, further X-ray observations, including detailed 
spectral variability study and high resolution spectroscopy 
such as obtainable with gratings on Chandra/XMM-Newton and
the calorimeter on Astro-E2 are needed. After characterizing the 
physical nature of the object, one can answer the most important question: 
''Why such AGNs are so rare? Is it just an extreme case 
of continuous population of type 1 AGNs with warm absorber or is there 
something essentially different about Kaz 102?''.


  This work has been partially supported by the NASA Grant NAG-10875 
(LTSA) to TM. The authors thank Tiziana Di Matteo for useful discussions 
on the ADAF possibility and kindly calculating a tailored ADAF SED for 
this paper. The authors are also thankful to the referee,  Kazushi Iwasawa, 
for various useful comments and pointing to many important 
references.

\end{document}